# Tuning Ultra-Narrow Direct Bandgap in $\alpha$-Sn Nanocrystals: A CMOS-Compatible Approach for THz Applications

‡*Tiziano Bertoli[1], ‡Elena Stellino[2], Francesco Minati[3], Camilla Belloni[4], Giovanni Tomassucci[3], Emanuele Bosco[1], Silvano Battisti[1], Leonardo Puppulin[4], Demetrio Logoteta[1] Alessandro Nucara[3], Luisa Barba[5], Gaetano Campi[6], Naurang Lal Saini[3], Fabrizio Palma[1,7], Michele Back[4*], Pietro Riello[4] and Fernanda Irrera,[1,7]*

[1]DIET, Sapienza University of Rome, P.le Aldo Moro 5, 00185 Rome, Italy

[2]Department of SBAI, Sapienza University of Rome Via Antonio Scarpa 14, 00185 Rome, Italy

[3]Department of Physics, Sapienza University of Rome, P.le Aldo Moro 2, 00185 Rome, Italy

[4]Department of Molecular Sciences and Nanosystems, Università Ca' Foscari Venezia
Via Torino 155, Mestre-Venezia, Italy

[5]Institute of Crystallography, CNR, Sincrotrone Elettra, Strada Statale 14, Km 163.5, Area Science Park, Basovizza, 34149 Trieste, Italy

[6]Institute of Crystallography, CNR, Via Salaria Km 29.300, Monterotondo, 00015 Roma, Italy

[7]Research Center for Nanotechnology for Engineering of Sapienza (CNIS), Sapienza University of Rome, Piazzale Aldo Moro 5, Rome, 00185, Italy

KEYWORDS. α-Sn, Nanocrystals, Ultra-narrow bandgap, Direct bandgap, THz applications, CMOS compatible process.

ABSTRACT

α-Sn has recently been attracting significant interest due to its unique electronic properties. However, this allotrope of Sn is stable only below 13 °C and alternative options to the conventional stabilization by epitaxial growth on InSb are still a challenge. In this work, nanoparticles with inner α-Sn nanocrystals were synthesized on a Silicon substrate via a CMOS-compatible process through microwave irradiation. The nanoparticle morphology was characterized by Scanning Electron Microscopy and Atomic Force Microscopy, demonstrating the ability to control the nanoparticle size by a dewetting process combined with a coalescence process induced by the microwaves. Grazing Incidence X-Ray Diffraction analyses confirmed the stabilization of the α-Sn phase within a $SnO_2$ shell, while X-Ray Photoemission Spectroscopy measurements revealed the presence of a bandgap. Infrared transmission spectroscopy combined with a Tauc-plot extrapolation led to an estimate of the gap in the range from 64 to 137 meV range. Furthermore, the possibility to tune the bandgap by controlling the nanoparticle size, possibly leveraging weak quantum confinement effects, was demonstrated, unveiling the potential of α-Sn nanoparticles on Si for the development of CMOS-compatible THz devices.

INTRODUCTION

Group IV semiconductors have been attracting considerable attention due to their intrinsic potential for integrated technologies compatible with complementary metal oxide semiconductor (CMOS) based on Si [1-3]. After the pioneering work of He and Atwater in 1997 [4], in the last few years the possibility of inducing a direct bandgap in $Ge_{1-x}Sn_x$ thin films, nanowires and nanocrystals [1,5-12] has been widely investigated. Despite the large lattice differences between Si, Ge and Sn, recent advances in the synthetic procedures allowed for the stabilization of compounds with narrow direct bandgaps ($E_g$) down to ~0.3 eV [13] (corresponding to a wavelength of about 4 µm) by closing the bandgap of Si and Ge alloyed with Sn . Less attention has been paid to the opposite situation, i.e. opening an ultranarrow bandgap in $\alpha$-Sn ($E_g$<250 meV), which could be exploited to bridge the THz gap [14], with potential applications in a variety of fields, ranging from the next generation mobile telecommunication system (5G, 6G) [15] to quantum cybersecurity [16] and astrophysics [17].

At ambient pressure, tin  exists in two main polymorphic structures: the thermodynamically stable tetragonal $\beta$-Sn (space group *I41/amd*) and the diamond-like cubic $\alpha$-Sn (space group *Fd-3m*) stable below 13 °C [18-24]. $\beta$-Sn is a metal while the electronic structure of $\alpha$-Sn is characterized by an inverted band, order consisting of an upper inverted light hole (iLH) $\Gamma_8^+$ conduction band and a lower heavy-hole (HH) $\Gamma_8^+$ valence band [25,26]. As a result, $\alpha$-Sn is semimetallic [22,27] and hosts topologically protected Dirac states [28-31].  This unique electronic structure potentially allows to tune the bandgap in the range of a few hundred meV, an energy window usually unavailable. However, the stabilization and control of the $\alpha$-Sn at operating temperatures is still a critical aspect for applications in modern electronics. Numerous studies have demonstrated stabilization at room temperature of $\alpha$-Sn films grown by molecular beam epitaxy (MBE) on substrates with diamond cubic cell structures and lattice constant very close to that of $\alpha$-Sn, such as InSb, CdTe, and GaAs. It is supposed that the stabilization of the $\alpha$-phase occurs thanks to the layer-by-layer epitaxial growth on these types of substrates [29,30,32,33]. Other authors reported the stabilization of $\alpha$-Sn on silicon substrates covered by a bilayer composed by thin Ge film and a thin Sn film on top, after rapid thermal annealing and quenching. The observed diffusion of approximately 2-4% Ge atoms in $\alpha$-Sn was hypothesized to trigger the $\alpha$-Sn stabilization by introducing mechanical stress [34,35]. However, among the strategies proposed for the stabilization of the $\alpha$-Sn, the effect of size is still debated. Indeed, some authors demonstrated the stabilization of the cubic $\alpha$-Sn phase at small size [36,37] in agreement with the general rule of the stabilization of the high symmetry phases, as already demonstrated for materials such as CdSe, $ZrO_2$, $Ga_2O_3$, $BaTiO_3$, $Al_2O_3$ [38-41], while others provided evidence of the opposite situation, evidencing the stabilization of the tetragonal $\beta$-Sn allotrope when the size decreases [42].

Despite the extensive investigation of the physical properties of $\alpha$-Sn grown on InSb and its recent stabilization on Si wafers through the use of Ge as a template and stabilizer [34,35], the control of the bandgap in an ultranarrow energy range and the development of a CMOS-compatible process able to integrate $\alpha$-Sn with the Si technology, by avoiding high temperatures and chemical processes, are desirable and still a huge challenge.

In this study, we report the controlled synthesis of Sn NPs and the evidence of the stabilization of inner $\alpha$-Sn nanocrystals with ultra-narrow direct bandgap controlled by the NP size. The NPs were synthesized on silicon, using a CMOS-compatible process patented by the Sapienza University, involving a baking step up to 400°C followed by microwave (MW) irradiation [43,44]. The morphological and structural investigation by means of Scanning Electron Microscopy (SEM), Atomic Force Microscopy (AFM) and Grazing Incident X-Ray Diffraction (GIXRD) allows to confirm the ability to control the NPs size and stabilize the $\alpha$-phase of Sn. Moreover, X-ray Photoelectron Spectroscopy (XPS) and Fourier-transform infrared spectroscopy (FTIR) investigation suggest the key role of the $SnO_2$ shell in the stabilization of $\alpha$-Sn and the possibility of tuning the absorption edge in the THz range by controlling the NP size.

EXPERIMENTAL RESULTS AND DISCUSSION

**Morphological and structural analysis: dewetting process and α-Sn identification**

As summarized in Figure 1, the process employed to stabilize α-Sn on the top of Si wafers consists of four steps: (i) deposition of an Sn film by physical vapor deposition (PVD), (ii) thermal treatment (baking) up to 400 °C followed by a (iii) cooling process until 200 °C and finally (iv) a MW irradiation in Ar atmosphere. Details regarding the process are reported in the Methods and Experimental Details section.

In order to evaluate the possibility of controlling the size and potentially the optical properties of the Sn nanostructures stabilized, Sn films of three different thicknesses, 4, 5 and 9 nm, were deposited (labelled as Sn4, Sn5 and Sn9, respectively) and subsequently treated by means of a two-step process consisting of a conventional thermal treatment (baking and cooling) and a MW irradiation (the samples that underwent all the fabrication steps are labelled as Snx_MW, with x=4, 5 or 9, while the samples that underwent all the steps except MW irradiation are labelled as Snx_bak).

To estimate the effective size of the nanoparticles, SEM analysis was conducted on the three different MW-irradiated samples (Figure 2a-c). The evaluation of the diameter distribution, conducted on a population of more than 6000 NPs for each sample (Figure 2d-f) provides an average value of 14, 17 and 20 nm for the samples Sn4, Sn5 and Sn9, respectively. The AFM analysis performed on the Sn4_MW and Sn5_MW samples (Figure 2g and h, respectively) confirms the average sizes of the NPs, also allowing an evaluation of the NP height with respect to the substrate (Figure S1). As shown in Figure 2i, there is a nearly linear relationship between the average NP diameter and the Sn film thickness, which suggests the possibility to finely control the nanosphere size. The dewetting of metal thin films has been modelled considering two different processes: (1) the hole nucleation and growth, typically encountered in the case of polycrystalline metal films [45-47], and (2) the spinodal dewetting [48,49]. The former predicts a linear behaviour between the NPs size ($d$) and the film thickness ($t$) while the latter is described by a $d \propto t^{5/3}$ power-law relationship. Based on the linear trend observed for $d$ *versus* $t$, a hole nucleation and growth mechanism could be suggested, but the extrapolation of the fitting straight line inconsistently predicts NPs with a diameter of about 10 nm in the limit of vanishing t. This can be explained considering that the process consists of a subsequential combination of thermal, cooling and MW irradiation treatments that cannot be described by a single dewetting process. In this view, to evaluate the effect of the single steps, for one of the samples (Sn5_bak), the process was stopped after the baking, letting the temperature reach the room value. The SEM and AFM inspections performed at the end of the baking/cooling step highlighted the presence of tiny NPs of about 8 nm in place of the uniform Sn film (Figure S2). Based on the large change in the size with respect to the corresponding sample treated by means of the MW, it is clear that the MW irradiation provides an appreciable size increase. This is likely due to the temperature-induced coalescence of nanoparticles. Indeed, it is well known that MW irradiation induces extremely rapid heating and cooling [50] resulting into an almost instantaneous change of the surface temperature from approximately 200 to 270°C and vice-versa (see Figure 2b). The presence of neck structures between the NPs detected by means of high-resolution AFM investigations (Figure S3) supports the coalescence between the tiny NPs formed during the baking step as the main process for the formation of larger NPs during the MW irradiation.

It is worth emphasizing that the process is independent of the kind of Si wafer employed, as the same behaviour is observed on heavily doped, lightly doped and intrinsic substrates, as well as on Si substrates covered by $SiO_2$ or $SiO_x$ films with thicknesses up to 70 nm.

In order to investigate the crystalline structure of the NPs grown on silicon substrate, Grazing Incidence X-ray Diffraction (GIXRD) measurements were carried out orienting the substrate along the (100) crystallographic direction and using x-ray synchrotron source at ELETTRA (12.4 keV). The diffraction images shown in Figure 3a reveal several distinct diffraction peaks. In panel (a), we present the GIXRD images of the bare Si substrate, as well as the Sn5_MW sample treated under microwave irradiation, which contains Sn nanoparticles. The Si (100) film clearly exhibits the main Bragg reflections. In contrast, Sn5_MW sample displays a significant number of experimental resolution-limited peaks. These additional peaks were indexed using the GIDVis software package [51] and were identified as reflections from Sn nanoparticles with cubic *Fd-3m* symmetry and lattice parameter of a=b=c=6.4892 Å. Therefore, the diffraction results provide clear evidence of an *α*-Sn phase in the samples treated under microwave with no evidence of β-Sn phase. In addition, the observed diffraction peaks indicate that the *α*-Sn phase of nanoparticles is a crystalline phase oriented along multiple directions, confirming a successful deposition and crystallization of Sn nanoparticles on the Si substrate. A few reflections corresponding to additional phases, such as $SnO_2$, are also present. It is worth mentioning that the α-Sn phase was reproducibly observed in samples grown using similar conditions (see, e.g., Figure S4 in the Supporting Information).

**Spectroscopic investigation: α-Sn stabilization mechanism and ultranarrow bandgap tuning**

XPS measurements were performed to gain a further understanding of the nature of the nanostructures and how the different treatments at different steps of the process affect them. In particular, the Sn5 sample at the different steps of the process (Sn5, Sn5_bak and Sn5_MW) was analysed. Firstly, survey spectra allowed to identify the core orbitals and the atomic species at the surface, as labelled in Figure 4a. Considering the surface sensitivity of the photoemission process, the presence of Si signals from the substrate can be justified owing to the intrinsically inhomogeneous covering of nanoparticles, while the occurrence of O and C is usually associated with adventitious impurities and/or the formation of chemically bonded states during the synthesis or the exposure to the atmosphere. Interestingly, the MW-irradiated sample exhibits a rather intense signal from Si, C and O orbitals, suggesting that the MW treatment may favour clustering and/or coalescence processes, leaving a larger portion of the substrate exposed to the probing x-ray photons, in agreement with the SEM and AFM investigations. To assess the possible role of the Si substrate in the stabilization of the $\alpha$-phase, XPS spectra of Si *2p* were measured (Figure 4b). The two broad spectral features around 100 eV and 103 eV indicate, respectively, the presence of $Si^0$ and $Si^{3+}/Si^{4+}$ species, suggesting the coexistence of different Si oxides at the surface of the Si substrate. However, none of the relevant spectral feature changes across the samples series, leading to the conclusion that neither heat treatment nor MW irradiation significantly affect the interaction between the substrate and the NPs, ruling out a major role of Si diffusion in the stabilization of the $\alpha$-phase as previously suggested [44]. On the other hand, the analysis of the Sn core levels offered valuable insights into the effects of the different treatments. We performed a peak-fitting analysis on Sn *3d* spectra of all samples modelling their spectral shape with three doublets, comprising the Sn *3d$_{3/2}$* and Sn *3d$_{5/2}$* contributions, separated by a spin-orbit split of 8.45 eV and reported in Figure 4c as $u_i$ and $v_i$ with $i$=1,2,3. Specifically, doublet 1 accounts for metallic $Sn^0$ contributions, which are likely to originate from the metallic core, while doublets 2-3 take into account $Sn^{2+}$-$Sn^{4+}$ contributions, aiming to model the Sn oxides SnO and $SnO_2$ which possibly constitute the NPs shell. Two additional singlet contributions ($p_1$ and $p_2$) have been included to model two plasmon peaks, allowing a proper background subtraction as well.

Sn5 sample exhibits a similar amount of SnO and $SnO_2$ (486.5 eV and 487.4 eV), together with a much smaller signal of metallic Sn (485.07 eV) possibly owing to the limited penetration depth of XPS in the NP core-shell geometry. While these ratios agree with previous reports [52], the spectral shape of Sn5_bak sample appears rather different. Firstly, the contribution of $SnO_2$ is almost suppressed while the metallic $Sn^0$ contribution has a much larger intensity ratio, allowing us to identify its binding energy at 484.99 eV with less uncertainty respect with film Sn5 sample. In order to rationalize this effect, we may assume for the as-grown Sn5 sample an oxidation gradient Sn-SnO-$SnO_2$ from the metallic core to the outer shell layers, as proposed in [52,53]. Experimental and theoretical evidence pointed out that the oxidation process during continuous heating mostly yields SnO in the considered temperature range [54,55] (even if the NP size can sensitively change the oxidation behaviour [56]). Hence, we may expect the outer $SnO_2$ shell to be disrupted by the heat treatment, leaving the NP with a shell mostly composed of SnO. As for the more intense $Sn^0$ signal from the metallic core, we can consider the possibility that such process induces an average thinner oxide layer when the dewetting process leads to the NPs formation.

After MW irradiation, the spectral shape of Sn $3d$ XPS spectra sustains a further considerable change. Surprisingly, in Sn5_MW sample, the larger spectral contribution comes from $Sn^{4+}$ peak, while the intermediate SnO oxide ratio appears strongly reduced, suggesting that the MW irradiation favours the stabilization of $SnO_2$ as NPs shell. However, we also observe a clear binding energy shift for both $Sn^0$ (485.15 eV) and $Sn^{4+}$ (487.75 eV) contributions. The higher binding energy of the Sn contribution can be interpreted as a signature of the $\alpha$-Sn phase stabilization in the NPs core, as previous XPS studies [57,58] indicate an energy shift of ~0.3 eV with respect to $\beta$-Sn phase. Here, the observed energy shift appears ~ 0.2 eV and the above interpretation is justified considering lower energy resolution in the present data. On the other hand, we can interpret the peculiar behaviour of $Sn^{4+}$ as a result of the possible stabilization of a crystalline phase of $SnO_2$, as the observed energy shift is compatible with the reported Sn *3d* XPS spectra of amorphous and crystalline $SnO_2$ [59]. Earlier structural analyses on Sn nanoparticles report an amorphous $SnO$-$SnO_2$ shell for as-grown samples [52,60], while the heating process mostly stabilizes the crystalline phase of SnO oxide in its early stages (200°C-400°C), a mixture of $SnO$-$SnO_2$ in the range 500°C-700°C and finally only crystalline $SnO_2$ is detected above 800°C [61,62]. Even if the oxidation process can be strongly dependent on the NPs size [56], it is apparent the crystalline phase of $SnO_2$ mostly occurs at high temperatures. In this context, the MW irradiation may act as an intense annealing process locally delivering an energy corresponding to an annealing treatment at much higher temperatures than the one reported in Figure 1 (~260°C). The crystallization process of $SnO_2$ shell has already been documented for both high-temperature annealing and microwave treatment accompanied by an increase NPs size [62,63].

Based on the information obtained so far, a possible mechanism for the stabilization of the $\alpha$-Sn can be proposed. As evidenced by the XPS analysis, the metallic transition of the Sn core from the $\beta$- to the $\alpha$-phase is accompanied by the complete transition from a major SnO character to a complete $SnO_2$ of the shell. The final stabilization of $\alpha$-Sn core and the $SnO_2$ shell after the MW treatment is also confirmed by the GIXRD analysis. When the SnO shell transforms into $SnO_2$, there is a volume contraction, which can induce tensile stress on the Sn core, because the shell shrinks and tends to pull the core. Such tensile stress, such as that induced by shell contraction, could favour the stabilization of the α-phase due to its larger volume compared to the β-phase (about 27% volume expansion [19]) relaxing the energy [64]. On the other hand, this is in agreement also with the observed tensile stresses on other oxidized metal NPs [65-67].

Finally, Fig. 3d shows the XPS valence spectra of all samples. According to previous photoemission studies [52,68], we labelled three energy regions on the basis of the Sn phase which provides the majority of the spectral weight. The electronic states near the Fermi level (up to a few hundreds of meV) belong to metallic Sn, while, approximately in the range 1-4 eV, the density of states is mostly attributed to Sn *5s* and Sn *5p* hybridized orbitals of SnO, and finally the feature appearing beyond 5 eV can be associated with O *2p* states of $SnO_2$. The behaviour of O *2p* offers a further heuristic argument to support the $SnO_2$ crystallization scenario in agreement with the GIXRD analysis, since it appears as a very broad feature around 6 eV in the Sn5 sample, while in Sn5_MW sample its profile tends to sharpen and to shift around 7.5 eV. On the other hand, in Sn5_bak the position of this feature is apparently aligned with the one of Sn5_MW, but it is strongly suppressed, suggesting an early $SnO_2$ crystallization process in the reported heating range coexisting with SnO. We observe by direct inspection that a small gap opens in MW sample, while the heat treatment appears to enhance the density of states at the Fermi level in Sn5_bak sample, which is possibly at the origin of the strong plasmon signal in the corresponding Sn *3d* spectrum. In the inset of Fig. 3d, we can closely observe the suppression of the states near the Fermi level in the Sn5_MW sample with respect to the pristine Sn5 film, confirming the occurrence of a gap opening. As the range of suppressed states only involves metallic Sn, we may address the gap opening as a further effect of the strain on the metallic NPs core or a quantum confinement effect. In normal conditions, $\alpha$-Sn exhibits a semimetal electronic structure, therefore it is possible that the very few electronic states near the Fermi level are highly susceptible to external perturbations. Interestingly, former studies on strained $\alpha$-Sn grown on InSb proved that upon strain application a topological gap of ~50 meV opens [69], and also its non-trivial topological properties have been investigated under compressive/tensile strain [70], opening the pathway for strain-engineered spintronic devices [34]. On the other hand, it is well known that quantum-confined structures could also experience a strong change in the optical behaviour due to the quantum confinement effect. However, to the best of our knowledge, there is no experimental report about the effect of size on the optical properties of $\alpha$-Sn nanocrystals.

The optical properties of the samples were investigated using FT-IR spectroscopy collecting transmission signals in the Mid-InfraRed (MIR) and Far-InfraRed (FIR) ranges and calculating the optical density ($OD$) as $OD = -log_{10}(I_{sample}/I_{bare})$, where $I_{sample}$ is the intensity of the radiation transmitted by the sample and $I_{bare}$ is the intensity of the radiation transmitted by the bare silicon substrate.

The $OD$ spectrum of all the samples subjected to MW irradiation displays a sigmoidal profile indicating the onset of inter-band electronic transitions in the FIR. Conversely, the spectrum of the non-irradiated samples after the baking step do not show any significant increase in the optical density signal in the explored energy range (see, as an example, the comparison between the $OD$ profiles of the Sn5_bak and Sn5_MW samples in Figure 5a) suggesting this transition to be related to the $\alpha$-phase. This result is consistent with both the GIXRD and XPS analysis confirming the stabilization of the $\alpha$-Sn only after the MW irradiation and thus indicating the MW irradiation as the key step for the stabilization of the semiconducting Sn phase.

The well-known Tauc-plot method [71-74] was employed to estimate the energy of the band gap $E_g$ of the semiconducting samples using the relation $(OD \cdot h\nu)^2 = B(h\nu - E_g)$. The $E_g$ value for each sample was then obtained by fitting the linear part of the curve in an appropriate range around the inflexion point and extrapolating the energy at $(OD \cdot h\nu)^2 = 0$, as shown in Figure 5b. Bandgap values of 64±2, 79±2 and 137±4 meV were estimated for the Sn9_MW, Sn5_MW and Sn4_MW samples respectively, which demonstrates a bandgap widening as the NP size decreases. The corresponding frequency window spanned by the samples is 15 to 35 THz (see the right panel of Figure 5c).

The dependence of the bandgap on the NPs size suggests attributing the band gap opening to quantum confinement (QC) effects. For a weak-confined electron-hole pair (exciton state) in spherical quantum dots (QDs), the bandgap energy can be expressed, with respect to the bandgap energy of the bulk as:

$$E_{QD} = E_{g,bulk} + \frac{\hbar^2\pi^2}{2R^2}\left(\frac{1}{m_e^*} + \frac{1}{m_h^*}\right) - \frac{1.786e^2}{\epsilon R}$$

where $R$ is the QD radius, $m_e^*$ and $m_h^*$ are the electron and hole effective masses, respectively, and $\epsilon$ is the dielectric constant. In the present case, $R$ corresponds to the radius of the $\alpha$-Sn core of the NPs. For analogous systems, it has been reported that the oxide shell thickness around the metallic Sn core is constant and approximately equal to 4.6 nm [60]. We assume a similar morphology for the MW-irradiated NPs and estimate the $\alpha$-Sn core radius by subtracting 4.6 nm from the radius of the NPs. When the size of the NP core is comparable to the Bohr exciton radius ($a_B^*$), the attractive Coulomb term becomes significant, leading to a decrease in the bandgap energy with respect to the QC effect. The Coulomb interaction strongly depends on the dielectric constant. In the case of $\alpha$-Sn, $\epsilon=24$ has been previously used to evaluate a Bohr exciton radius of 12.56 nm [37]. However, when the QDs are embedded in a material with a smaller dielectric constant, the screening effect is reduced and the Coulomb interaction is enhanced [75]. This effect can be modelled by considering an effective dielectric constant, with a value intermediate between that of the oxide shell and of the core.

In order to evaluate the hole and electron effective masses, the band structure of $\alpha$-Sn was calculated by means of DFT simulations. The electronic structure obtained (Figure S5) agrees with previous reports [76]. The electron effective mass in the lowest conduction band, was found to be isotropic and equal to $\approx 0.04\ m_0$, where $m_0$ is the electron rest mass. On the contrary, the hole effective mass in the highest valence band is markedly anisotropic and an average value of $\approx 0.2\ m_0$, obtained as the harmonic mean of the tensor principal values, was assumed. By using these values and an effective dielectric constant of 14, a Bohr exciton radius $a_B^* = 10.6$ nm was estimated. The plot of $E_{QD}$ as a function of $R$ is shown in Figure 5c. The experimental bandgap values gather around the curve, indicating that quantum confinement could actually play a major role in determining the bandgap behavior.

CONCLUSIONS

In an upside-down view with respect to the closing of the Si and Ge bandgap conventionally employed to redshift in the mid-IR range, in this work we demonstrated the possibility to open and control an ultranarrow band gap in $\alpha$-Sn.

Sn NPs with $\alpha$-phase cores were successfully synthesized and stabilized on silicon substrates using a CMOS-compatible process based on baking of uniform tin films and subsequent irradiation of the NPs by microwaves. The process enables precise control over the size of the nanoparticles, which can be valuable for the future engineering of devices based on this technology.

Synchrotron radiation GIXRD demonstrated the presence of the $\alpha$-phase, further confirmed by XPS analysis. Optical characterization performed with FTIR spectroscopy revealed the presence of an absorption profile in the far infrared spectrum in all the samples irradiated with the microwaves, in agreement with the presence of the $\alpha$-phase stabilized by the MW irradiation. Direct transitions with energy gap values ranging between 64 and 137 meV (corresponding to threshold frequencies between 15 and 34 THz) were demonstrated. The stabilization of the $\alpha$-phase at room temperature appears to be highly reproducible and can be explained by considering the effect of the oxide shell that induces a tensile strain in the metal core of Sn, promoting the stabilization of the expanded $\alpha$-Sn phase. On the other hand, the relationship between the optical bandgap and the Sn NPs size is consistent with a weak quantum confinement effect, which would open the possibility of significantly expanding the covered frequency range by better controlling the NPs size and shape. The full compatibility of the fabrication process with CMOS technology provides the opportunity of developing integrated solutions and can pave the way towards significant advancements in THz technology.

METHODS AND EXPERIMENTAL DETAILS

**Sn Thin Film Deposition and α-Sn NCs Stabilization Process**

Sn thin films were deposited by PVD on Si wafer with a Balzer 510 evaporation system, with a vacuum of $2 \cdot 10^{-6}$ mbar and a deposition rateo of 0.70 Å/s. Then, the Sn films deposited on the Si wafers were placed in a CVD chamber, where they underwent a baking step in which the temperature was raised to 400°C with a ramp rate of 3800°C/h by a heating plate placed in contact with the substrate. After, the sample is allowed to cool freely within the chamber until it reaches 200°C. Finally, a MW irradiation with a power of 450 W for 3 minutes in an Ar atmosphere (pressure = 2 mbar, flow = 20 sccm) was employed. MWs were generated by a 2.45 GHz Alter Ti Series microwave source. The surface temperature profile during the whole fabrication process is measured by an infrared pyrometer (METIS M318 Sensortherm) as summarized in Figure 1b.

**Morphological Investigation**

The morphological investigation was performed by means of SEM (Zeiss model AURIGA) working between 3 and 8 kV, using InLens detector to collect secondary electrons in high vacuum condition. The distributions were analysed more than 6000 NPs for each sample.

AFM topography images were acquired in tapping mode using a Dimension Icon atomic force microscope (Bruker, Germany) in ambient conditions. We used a sharp AFM tip with 1 nm end radius fabricated on a cantilever with spring constant and resonant frequency of 0.12 N/nm and 100 kHz, respectively (Peakforce-HIRS-SSB, Bruker, Germany). Images of the sample surface were collected at scan sizes of 900, 500 and 200 $nm^2$ and scan rate 0.5 Hz.

**Structural Measurements: Synchrotron radiation GIXRD**

Grazing Incidence X-ray Diffraction (GIXRD) measurements were performed using x-ray synchrotron source at ELETTRA, Trieste at the crystallographic XRD1 beamline at the ELETTRA synchrotron facility in Trieste. The beam energy was set to 12.4 keV using a vertical collimating mirror and a double-crystal Si(111) monochromator. Initially, the samples were aligned along the vertical z-axis, and the grazing incidence angle was determined experimentally. The measurements were performed by rotating the sample around the z-axis, which is perpendicular to the film surface, covering an angle of 180 degrees with an exposure time of 20 seconds. A beam size of 0.2 x 0.2 mm² was used. Diffraction images were collected using a Dectris Pilatus 2M detector, which has 1475 x 1679 pixels with an area of 172 x 172 μm². The sample-detector distance was set to 150 mm. Diffraction data were calibrated using a LaB$_6$ standard and analyzed with the GIDVis software package [51].

**Spectroscopic Investigation: XPS and FT-IR**

All the XPS measurements were carried out at room temperature, exploiting an in-house ultra-high vacuum setup equipped with a double Al-Mg anode SPECS XR50 x-ray source and a multichannel Omicron EA125 electron analyzer, operating at base pressure $< 2 \times 10^{-9}$ mbar. The Al Kα excitation line (hν=1486.6 eV) was employed to probe the electronic structure of a sample series.

The optical properties of the samples were investigated using FT-IR spectroscopy (Bruker IFS 66 v/s) with a custom-built setup in transmission configuration, placing the sample directly in front of the detector to minimize scattering effects due to the NP size and surface roughness of the substrate. Measurements were conducted in the Mid-InfraRed (MIR) and Far-InfraRed (FIR) ranges using a DTGS detector coupled with a KBr and a Multilayer Mylar beamsplitter respectively.

## Computational Methods

DFT simulations were performed in a plane-wave basis using the Quantum Espresso package [77]. The Heyd–Scuseria–Ernzerhof exchange-correlation functional [78] and fully-relativistic norm-conserving Vanderbilt pseudopotentials without non-linear core corrections were adopted. The plane-wave energy cutoff was set to 50 Ry and the first Brillouin zone was sampled using a 12x12x12 Monkhorst-Pack grid. The α-Sn primitive cell was relaxed without constraints until interatomic forces were lower than $10^{-5}$ Ry/Bohr and the band structure was subsequently computed via interpolation on a maximally localized Wannier function basis [79]. The effective mass tensors were obtained from the curvature of the highest valence band and of the lowest conduction band in Γ, evaluated within a second-order finite-difference approximation.

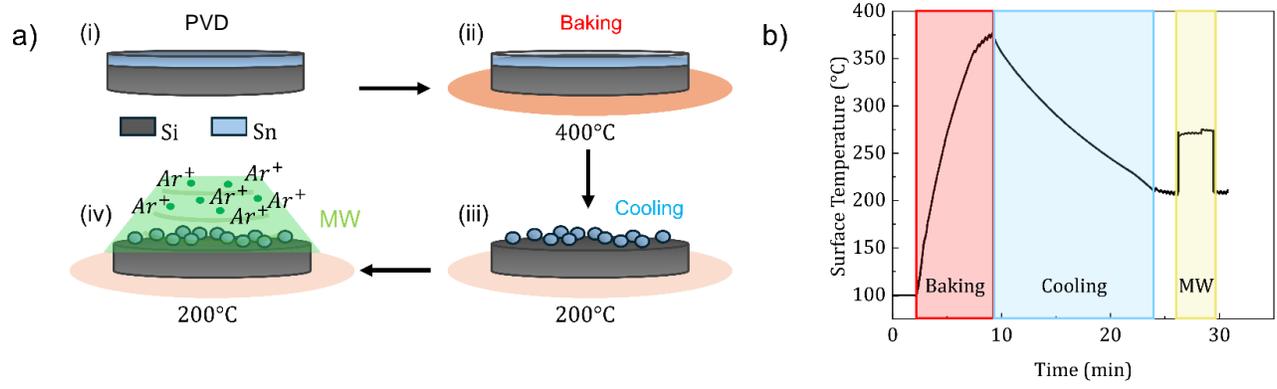

**Figure 1.** a) Process steps: (i) Sn film deposition by PVD, (ii) baking, (iii) cooling and (iv) MW irradiation. b) Surface temperature profile.

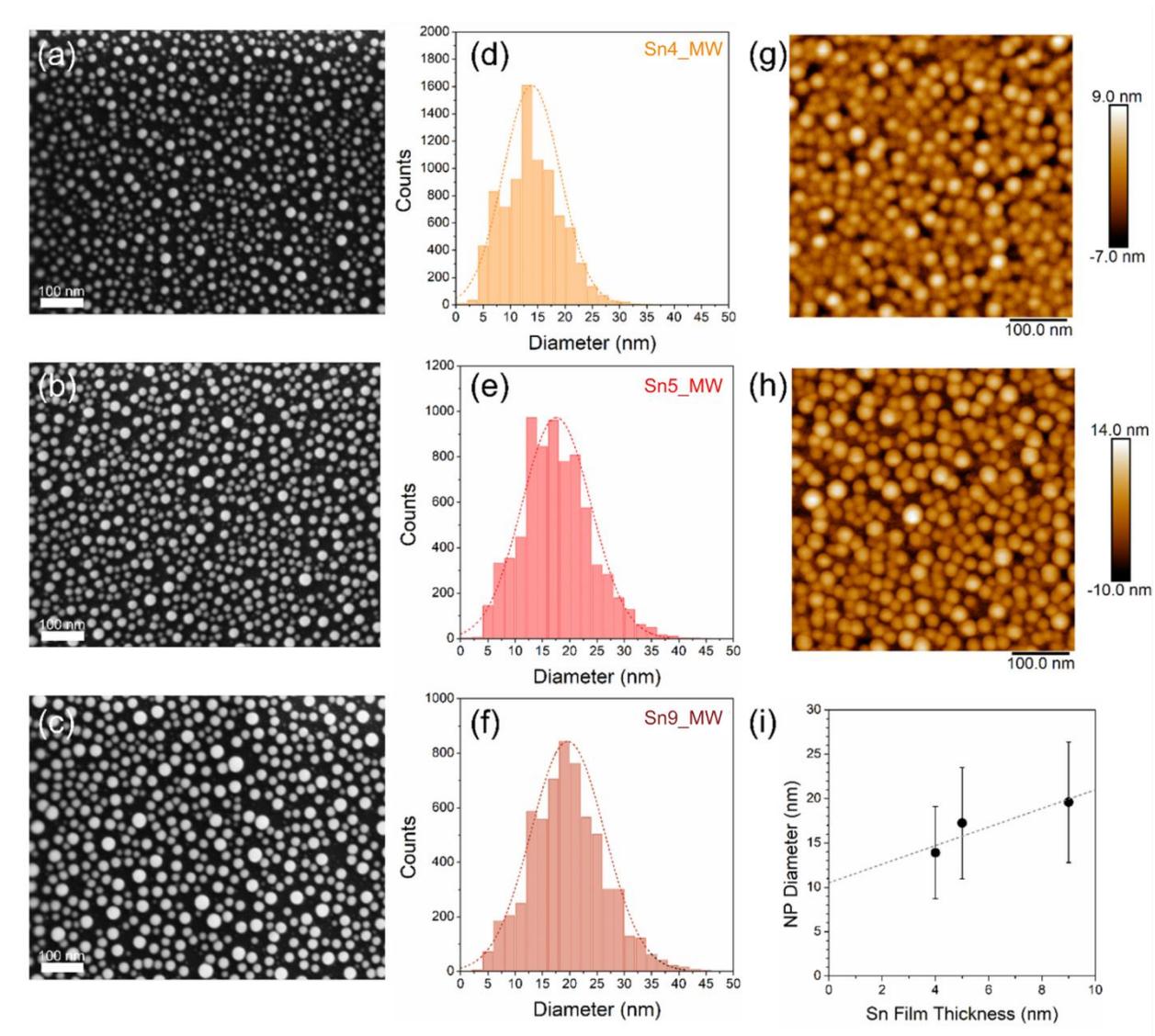

**Figure 2.** (a-c) SEM images of samples Sn4_MW, Sn5_MW and Sn9_MW along with the corresponding size distributions (d-f). AFM topography of (g) Sn4_MW and (h) Sn5_MW samples and (i) NPs diameter vs deposited Sn film thickness relationship.

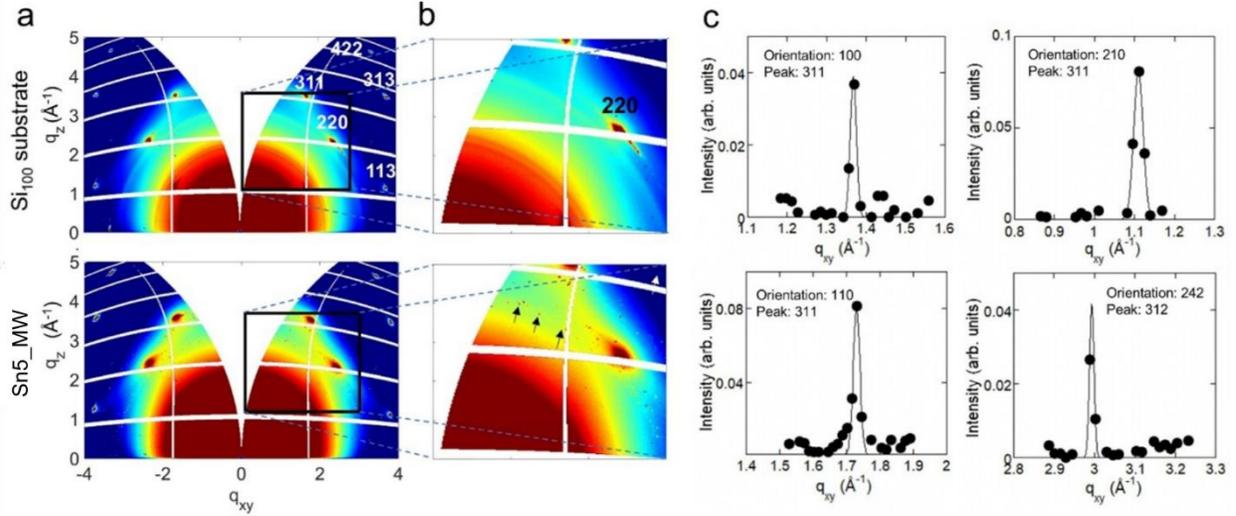

**Figure 3.** (a) GIXRD images of the Si(100) substrate and the Sn5_MW sample. (b) Magnified views around the 220 Bragg peak of Si(100), showing resolution-limited peaks associated with highly crystalline cubic α-Sn grains. (c) Background-subtracted GIXRD profiles integrated along the $q_z$ direction, highlighting reflections corresponding to the peaks marked by arrows in the magnified section (b). The *hkl* indices, along with their respective orientations, are indicated. For further comparison, Figure S4 in the Supporting Information includes GIXRD data on a second sample.

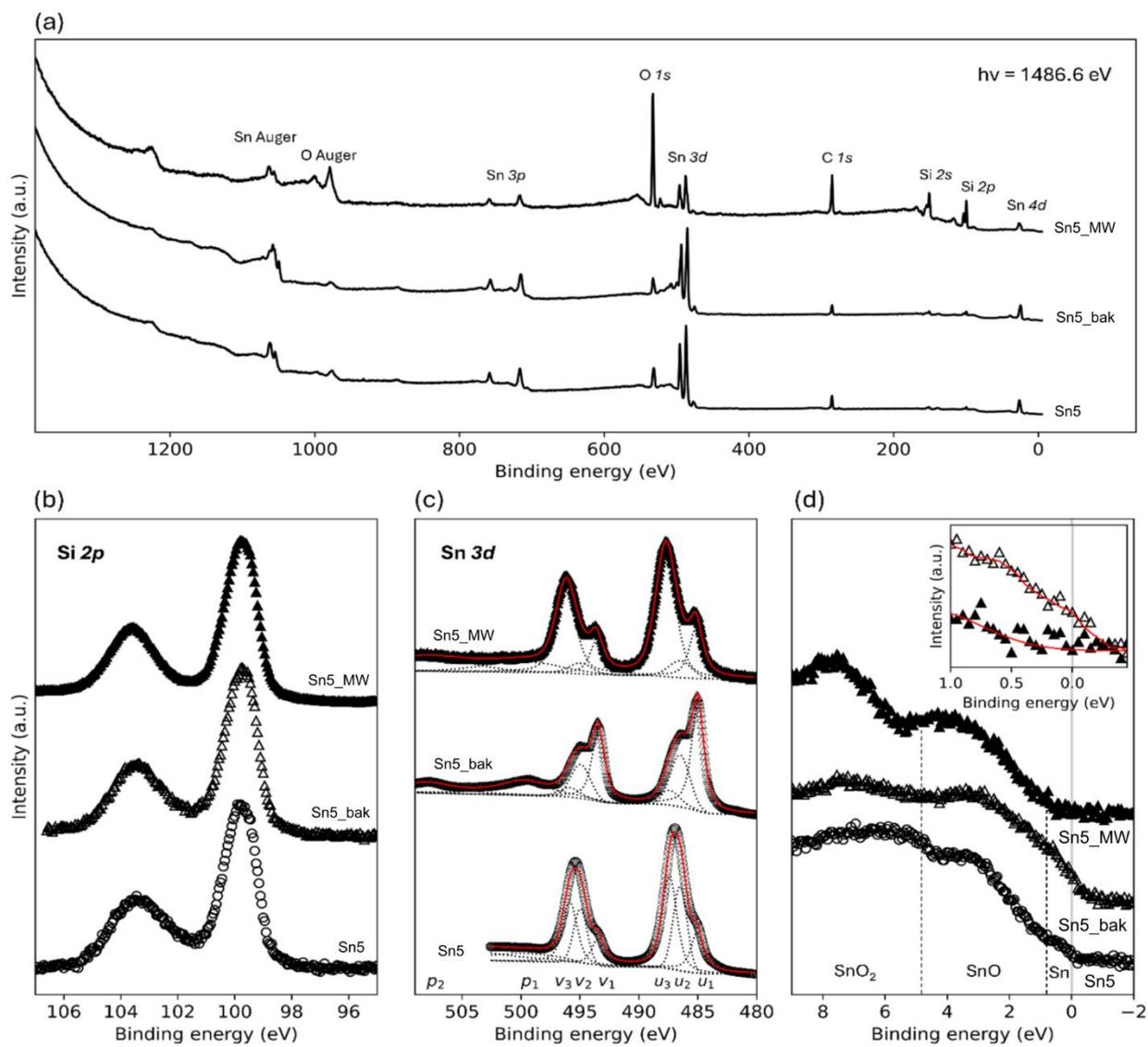

**Figure 4.** (a) XPS survey scans of baked (Sn5_bak) and microwave-treated (Sn5_MW) samples are shown together with the Sn5 reference sample. (b) Si *2p* XPS; (c) Sn *3d;* and (d) valence band spectra are shown for the samples at the three stages. The Sn *3d* XPS spectra are deconvoluted in three components (dotted lines) and assigned to Sn ($u_1,v_1$), SnO ($u_2,v_2$) and SnO$_2$ ($u_3,v_3$). The inset to the panel d) is a zoom-in of the valence states near the Fermi level of the Sn5_bak and microwave-treated Sn5_MW samples to highlight the non-metallic nature of the latter.

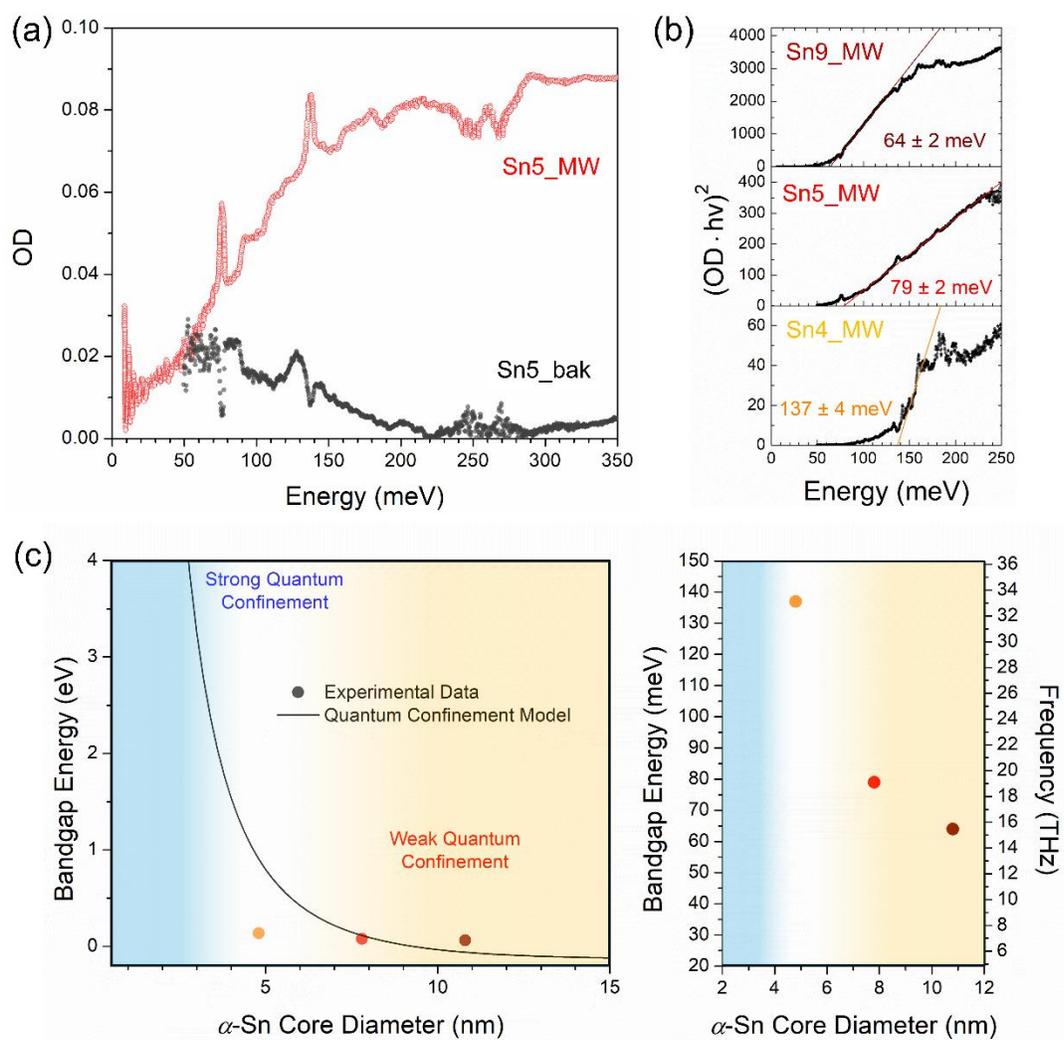

**Figure 5.** (a) Optical density of the absorption spectra of the Sn5_bak and Sn5_MW samples. (b) Tauc plot of the samples. (c) Bandgap energy as a function of the α-Sn core size along with the curve obtained from the quantum confinement model. An effective dielectric constant was used to take into account the presence of the outer $SnO_2$ shell. (d) Enlargement of panel (c), highlighting the frequency window spanned by the experimental points.

AUTHOR INFORMATION

**Corresponding Author**

*michele.back@unive.it

*fernanda.irrera@uniroma1.it

**Author Contributions**

The manuscript was written through contributions of all authors. All authors have given approval to the final version of the manuscript. ‡Tiziano Bertoli and ‡Elena Stellino contributed equally.

SUPPORTING INFORMATION

Additional experimental (SEM, AFM, XPS) and computational (DFT) data.

ACKNOWLEDGMENT

This study was carried out within the project "UltraNarrow Bandgap Engineering of alfa-Sn towards Mid-Infrared/THz Silicon Technology" and received funding from the European Union Next-GenerationEU - National Recovery and Resilience Plan (NRRP) – MISSION 4 COMPONENT 2, INVESTIMENT 1.1 Fondo per il Programma Nazionale di Ricerca e Progetti di Rilevante Interesse Nazionale (PRIN) – CUP N. H53D23004710006 (grant number 2022X2Y8SJ). This manuscript reflects only the authors' views and opinions, neither the European Union nor the European Commission can be considered responsible for them.

E.S. acknowledges funding from Project ECS 0000024 Rome Technopole, – CUP B83C22002820006, NRP Mission 4 Component 2 Investment 1.5, Funded by the European Union